\documentclass[structabstract]{aa}  
\usepackage{graphicx}
\usepackage{txfonts}
\usepackage{natbib} 

\begin{document}
   \title{{\it RHESSI} observations of long-duration flares \\ with long-lasting X-ray loop-top sources}

   \author{S. Ko{\l}oma\'nski \and T. Mrozek \and U. B\c{a}k-St\c{e}\'slicka}

   \institute{Astronomical Institute of the University of Wroc{\l}aw, Kopernika 11, 51-622 Wroc{\l}aw, Poland \\ 
   \email{(kolomans, mrozek, bak)@astro.uni.wroc.pl}}

   \date{Received ... ; accepted ...}


  \abstract
   {The \textit{Yohkoh}/HXT observations of Long Duration Events (LDEs) showed that the HXR emission ($14-23$~keV) is present for tens of minutes after flare maximum. Hence, some heating process is expected to exist during that time. The better energy resolution of \textit{RHESSI} compared to HXT allow us to analyse LDEs in more comprehensive way.}
   {We selected three LDEs observed by \textit{RHESSI} to answer the questions how long HXR emission can be present, where it is emitted, what is its nature and how much energy should be released to sustain the emission.}
   {We used \textit{RHESSI} data to reconstruct images of the selected flares with an energy resolution as good as 1 keV. Next we estimated physical parameters of HXR sources through imaging spectroscopy. The physical parameters obtained were then used to calculate the energy balance of the observed sources.}
   {We found that HXR thermal emission can be present for many hours after LDE flare maximum. The emission comes from large and hot loop-top sources. The total
   energy that must be released to sustain the emission of the sources is as high as $10^{31}\;{\rm erg}$.}
   {}

   \keywords{Sun: flares - Sun: X-rays, gamma rays}

   \titlerunning{{\it RHESSI} observations of long-duration flares}
   \maketitle


\section{Introduction}

A long duration event (LDE) is a solar flare characterized by a slow decrease in soft X-ray (SXR) thermal emission. This decrease may last from several hours to more than a day. Much insight into the nature of LDEs was made by ultraviolet and X-ray observations, during the \textit{Skylab}, \textit{SMM} (Solar Maximum Mission) and \textit{Yohkoh} space missions \citep[e.g.][]{sheeley1975, kahler1977, feldman1995, murnion1998, czaykowska1999, shibasaki2002, isobe2002}. One of the most important conclusions is that without the continuous energy input during the whole decay phase LDEs would decay much faster than it is observed.
 
Loop-top sources (LTSs) are remarkable SXR and HXR (hard X-rays) features of solar flares seen close to a flare loop apex. They form before flare maximum and in LDEs may last many hours \citep[e.g.][]{feldman1995, kolomanski2007a}. LTSs should be located close to the primary energy release site \citep[e.g.][]{kopp1976, shibata1999, hirose2001, karlicky2006}. Thus, they are a very promising tool for an analysis of an energy release during the decay phase. LTSs were first recorded on images taken from \textit{Skylab} and are commonly present in the \textit{Yohkoh}/SXT flare observations. Since the first observation it has become clear that the presence of an LTS during the whole flare decay-phase requires continuous energy release and some restriction mechanism efficiently preventing outflow of mass and energy from an LTS \citep[see][]{vorpahl1977}. Without meeting these two requirements loop-top sources would rapidly lose energy by radiative and conductive processes and would vanish. This result was later confirmed by an analysis based on \textit{RHESSI} and \textit{Yohkoh} data \citep[e.g.][]{jiang2006, kolomanski2007b}.

In the CSHKP \citep[e.g.][]{carmichael1963, hirayama1974, kopp1976, sturrock1966} model of a solar flare the energy release occurs above an LTS. This model and its numerous modifications \citep[][and references within]{krucker2008} successfully explain observational features of solar flares observed in different wavelengths during the impulsive phase. However, the authors investigating LDEs \citep[][]{nakajima1998, uchida1999, kundu2001, phillips2005} report observations inconsistent with the CSHKP concept. The main problems they had were related to non-thermal velocities of 20-50 km/s, long times needed to keep the inflow of a magnetic field into a reconnection site and a lack of downflows or upflows. Recently \citet{jiang2006} investigated heating and cooling processes of an LTS and found that there is a large amount of energy released during the decay phase of a short-duration solar flare. The authors suggest that in the case of an LDE the energy released during the decay phase may be even larger than the energy released during the impulsive phase. 

An analysis of energy release during the decay phase can give us precise constraints for numerical flare models. The most demanding for the models is long-lasting HXR emission of LTSs seen in LDE flares. Therefore such LTSs observed for many hours after a flare maximum are probably the most promising observational feature to set up the constraints. The sources should be very close to the energy release site and, that is the most important attribute, they put high requirements on the energy release rate. If a long-lasting HXR source is thermal then it must be continuously heated, because the characteristic radiative cooling time of hot (above 10 MK) and dense ($\approx10^{10}\;{\rm cm}^{-3}$) plasma is $\approx 1$~hour. If an HXR source is non-thermal then there should be a continuous acceleration of particles, to counteract fast thermalization of non-thermal electrons. Time-scale of the thermalization in the sense of momentum loss of electrons is about several seconds.

Hard X-ray LTSs were reported before with {\it Yohkoh}/HXT data, but with observations shorter than thermal soft X-ray LTSs. \citet[]{masuda1998} observed HXR sources up to $30$ minutes after flare maximum, with typical sizes $1-2$~arcmin. \citet[]{murnion1998} investigated two LDEs and concluded that HXR sources grow with time, with diameters $\approx20-45$~arcsec. In one of these flares the HXR source was visible up to 3 hours after flare maximum. \citet[]{kolomanski2007a, kolomanski2007b} analysed three LDEs using {\it Yohkoh} data, finding that HXR sources during the decay phase were well correlated with SXR LTSs and were observed (in $14-23$~keV range) for about 50 minutes after flare maximum. \citet[]{khan2006} reported the observation of thermal HXR source lasting more than 1 hour after flare maximum. In all these cases the HXR emission was observed close to and above the SXR emission source. Although {\it Yohkoh}/HXT data have provided useful information about LDEs, generally only L channel ($14-23$~keV) showed significant emission for any length of time. It was thus not possible to distinguish between the thermal and non-thermal nature of an LTS.

\citet[]{gallagher2002} observed evolution of an LTS using {\it RHESSI} and {\it TRACE} data. The HXR emission was observed in the $12-25$~keV range some 4 hours after the flare maximum and almost 11 hours in the $6-12$~keV range. The LTS was large and its altitude increased at a speed that gradually declined from $10$ to $1.7\;{\rm km}\:{\rm s}^{-1}$. The higher energy emission ($12-25$~keV) was located above the lower energy ones ($6-12$~keV and $3-6$~keV) and all HXR emission was located above the tops of the loops observed in the EUV range ({\it TRACE}~195~\AA). 

So far, most investigations of LDEs made with {\it RHESSI} have been restricted to a few wide (several keV) energy intervals. Recently \citet[]{sainthilaire2009} presented observations of a coronal source recorded by {\it RHESSI}. The source was observed for about 12 hours in energies up to 10~keV. Emission of the the source was purely thermal with maximal temperature of about 11~MK. Authors estimated a total energy (about $4\times10^{31}\;{\rm erg}$) which had to be deposited in the source to sustain its long lasting emission.

Here we present the investigation of LDEs observed by {\it RHESSI}. The analysis is made using {\it RHESSI} images reconstructed in very narrow ($1$~keV) energy intervals. {\it RHESSI} is very useful for analysing weak sources even with 1~keV energy resolution due to its high spectral resolution. Thus, it gives an opportunity to determine what is the nature of LTSs emission -- thermal or non-thermal. Using the images we estimate LTSs physical parameters through imaging spectroscopy. Then the parameters are used to calculate the energy balance of the observed sources and to find out how effective are the energy release and heating processes at the decay phase of an LDE.


\section{Observations}

Three LDEs, well observed by \textit{RHESSI} \citep[]{lin2002}, were selected for an analysis. We chose flares of significantly different power and with decay phases lasting more than 10 hours in \textit{GOES} $1-8$~\AA\,range. The flares occurred on: 30 July 2005 (hereafter the LDE1), 22 August 2005 (LDE2), 25 January 2007 (LDE3). Basic information about the LDEs is given in Table~\ref{tab1}. Our analysis is based on {\it RHESSI} data \citep[Reuven Ramaty High Energy Solar Spectroscopic Imager,][]{lin2002} supported with {\it SoHO}/EIT \citep[Extreme UV Imaging Telescope on-board Solar and Heliospheric Observatory,][]{delaboudiniere1995} and {\it GOES}/SEM \citep[Space Environment Monitor on-board Geostationary Operational Environmental Satellites,][]{donnelly1977} observations.


\begin{table*}
\begin{minipage}{\textwidth}
\caption{Basic information about the selected long-duration flares.} 
\label{tab1} 
\centering 
\begin{tabular}{c c c c c c c c} 
\hline\hline 
\noalign{\medskip}
Flare & Date &    \multicolumn{2}{c}{Location}                      & \multicolumn{2}{c}{SXR Flux Maximum\tablefootmark{3}} & Duration\tablefootmark{3} & Time Intervals Analysed \\ 
      &      & Active Region\tablefootmark{1} & Coordinates\tablefootmark{2} &     Time [UT]    &     Class         &    [h]   &       Beginning [UT] (Duration [min]) \\ 
\noalign{\medskip}
\hline 
\noalign{\medskip}
LDE1 & 30 July 2005 & 10792 & N10E59 & 06:36 & X1.3 & 11 & 10:20 (0.07), 11:30 (0.13), \\ 
     &              &       &        &       &      &    & 11:55 (0.13), 12:40 (0.27), \\
     &              &       &        &       &      &    & 13:34 (0.53), 14:15 (1.00), \\
     &              &       &        &       &      &    & 15:55 (4.00), 16:26 (4.33)  \\ 
\noalign{\medskip}
LDE2 & 22 August 2005 & 10798 & S10W52 & 01:34 & M2.7 & 11 & 01:35 (0.13), 01:55 (0.27), \\
     &                &       &        &       &      &    & 02:45 (2.00), 03:24 (1.00), \\
     &                &       &        &       &      &    & 04:22 (1.00), 04:46 (4.00), \\
     &                &       &        &       &      &    & 05:56 (8.00), 06:36 (2.00), \\
     &                &       &        &       &      &    & 07:29 (2.00), 08:08 (2.00)  \\     
\noalign{\medskip}
LDE3 & 25 January 2007 & 10940 & S07E90 & 07:15 & C6.3 & 17 & 07:03 (1.50), 07:55 (0.13), \\
     &                 &       &        &       &      &    & 08:12 (0.33), 08:32 (0.50), \\
     &                 &       &        &       &      &    & 09:30 (1.50), 09:50 (1.50), \\
     &                 &       &        &       &      &    & 10:10 (1.50), 11:00 (2.50), \\
     &                 &       &        &       &      &    & 11:22 (3.00), 11:47 (4.00), \\ 
     &                 &       &        &       &      &    & 12:30 (4.00), 12:56 (4.00)  \\     
\hline 
\end{tabular}
\tablefoottext{1}{NOAA active region number;}
\tablefoottext{2}{heliographic coordinates;}
\tablefoottext{3}{according to \textit{GOES}  $1-8$~\AA\,database}
\end{minipage}
\end{table*}

\subsection{On the {\it RHESSI} data}

{\it RHESSI} is a rotating Fourier imager with nine detectors made of pure germanium crystals \citep[]{lin2002}. The detectors record energy and arrival time for each detected HXR photon. For strong solar flares the number of coming photons easily reaches $10^5$ counts per second. As the number of counts rises the lifetime of a detector decreases. The pulse pile-up \citep[]{smith2002} is the main problem connected with this effect. It occurs when two or more photons arrive almost at the same time. In such a case the electronics of the detector recognize them as one photon with energy equal to the sum of these photons. When number of incoming photons is very high the pulse pile-up may severely influence measured fluxes.

The number of HXR photons decreases strongly with energy so problems with pulse pile-up occur mainly in the low-energy end of the range observed by {\it RHESSI}. Thus, to achieve the great dynamic range of the instrument, several systems to decrease the measured low-energy HXR photons were included. Among these systems there are attenuators which are the most important source of uncertainty. Influence of the attenuators on measured fluxes is quite well understood at present, but in practice the attenuator could seriously complicate the spectral analysis \citep[]{smith2002} especially in the lowest energies. 

For each orbit, \textit{RHESSI} passes through the radiation belts and South Atlantic Anomaly (SAA) which may have also a significant influence on the measured HXR fluxes. In our analysis we focused on the late phase of an LDE when the emission is extremely weak, so some care is needed in the analysis. However, the image spectroscopy we performed has an advantage in comparison to "standard" spectroscopy based on fluxes measured for the whole Sun. The background photons do not influence modulation profile for the rotating Fourier imager. Thus, an analysis of very weak fluxes can be done much easier than in the case of "standard" spectroscopy, especially for moments of passage through radiation belts when the background changes very quickly with time. 

\textit{RHESSI} light curves of the LDEs are shown in Figs.~\ref{fig20050730ligcur}, \ref{fig20050822ligcur} and \ref{fig20070125ligcur}. The effect of radiation belts and SAA passages can be seen on the light curves. Usually the SAA influence is observed as a rise of the signal observed several minutes before and after the passage. When satellite is in the SAA detectors are switched off. The influence of the radiation belts is visible as a slow oscillation of the signal. 


\section{Analysis}

\subsection{Imaging}

It is a great difficulty to reconstruct a {\it RHESSI} image when an emitting source is weak. Usually {\it RHESSI} image reconstruction is performed for detectors Nos. 3--6, 8 and 9. Depending on the reconstruction algorithm and weights chosen this set of detectors gives a spatial resolution about $7-9$~arcsec \citep[]{aschwanden2002}. However, we were not able to reconstruct any reliable image in a case of very late phase of LDE decay with the use of this standard set of grids. The weak signal was not the problem since taking integration times up to 8 minutes we were able to collect enough counts, i.e. about several thousand. 

The problem may be solved if we remember that when the source size is comparable to the resolution of a particular grid then the detector records very weak or no modulation of the signal \citep[]{hurford2002}. The resulting image obtained for this grid is then dominated by noise. Such grid added to the set used for reconstruction introduces only noise. In the worst case we may not get convergence of the reconstruction algorithm.

\begin{figure*}
\sidecaption
\includegraphics[width=12cm]{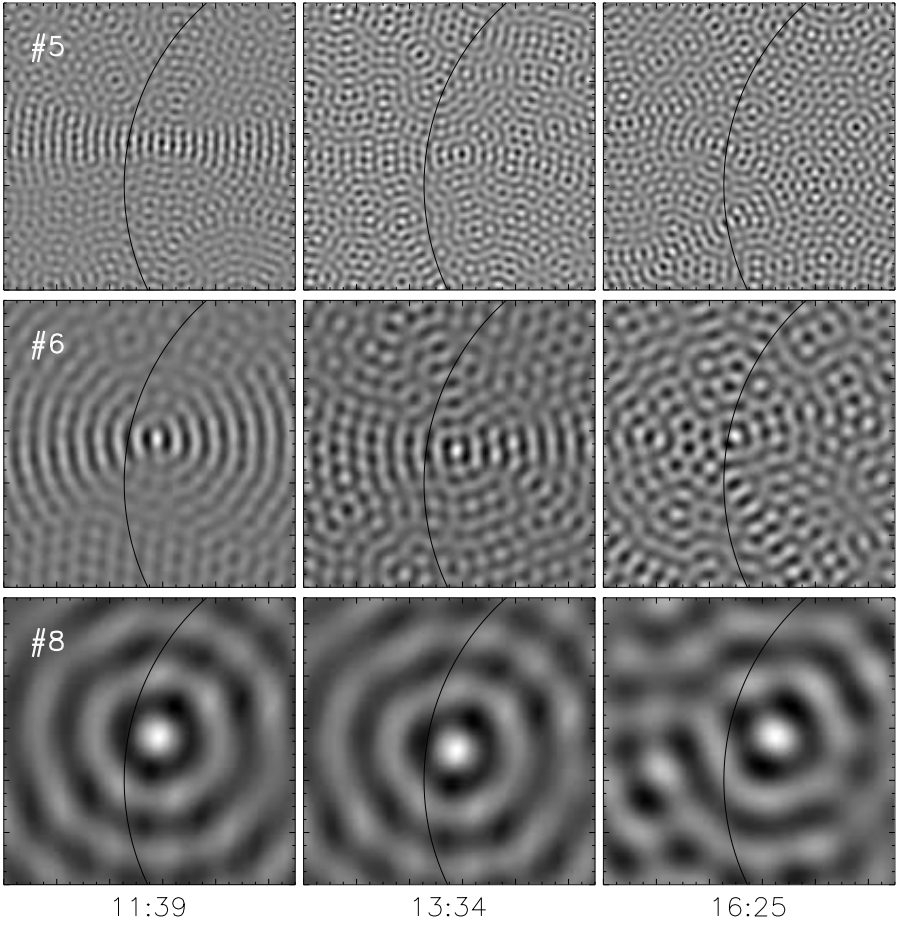}
\caption{The $6-7$~keV back-projected \textit{RHESSI} images reconstructed for selected grids (Nos. 5, 6 and 8) and three time intervals (the 30 July 2005 flare). 
When a source size is comparable to the resolution of a particular grid then image obtained for this grid is dominated by noise. Field of view of each image is 
$1024 \times 1024$~arcsec. The Sun limb is marked be a black line.}
\label{singledet}
\end{figure*}

We tried to solve this problem in the following way. First, we reconstructed images for single grids and a wide field of view (8~arcsec pixel, image size 256$\times$256 pixels) using the back-projection algorithm \citep[]{hurford2002}. Next, we used these images for determining which grid provides us with reliable images of the source. An example is presented in the Fig.~\ref{singledet}. For some time intervals there is only noise in the narrow grids, although count rates are high. Taking this into account we selected, for a given time interval, only those grids that showed a definite source in a single detector image. 

In the next step we used chosen set of grids to reconstruct images with PIXON algorithm \citep[][and references within]{puetter1999}. Since integration times are rather long (minutes) we used stacked modulations \footnote{http://sprg.ssl.berkeley.edu/$\sim$tohban/nuggets/ ?page=article\&article\_id=39}. Images obtained with this method were more "stable" than without it. "Stable" means that images in neighbouring time and energy intervals were similar. Moreover there were fewer images where we did not get convergence of the reconstruction algorithm. Energy resolution chosen for reconstruction was as narrow as possible, i.e. 1~keV. It enabled to obtain reliable fits to observed spectra.

\subsection{Geometrical properties -- LTS centroid location, altitude and size}

Having reconstructed set of images we were interested in obtaining some properties of observed sources. In each image we determined sources using 50\% of brightest pixel intensity isoline. In each case we observed sources related to coronal parts of flaring structure only (loop-top sources). There was no observed emission localised on the chromosphere level (footpoint sources) which is rather typical for the decay phase of an LDE.

We determined position of loop-top sources by finding their centroids. The centroids were estimated as centres of gravity of the emission within the 50\% intensity isoline (relative to the brightest pixel in the {\it RHESSI} image). The uncertainty of the source location estimation is caused by two agents. First agent is an uncertainty connected with spatial resolution of the instrument although using centroids we significantly improved the accuracy of source location estimation with regard to nominal. In our analysis this error was always less than 1.5 arcsec. More uncertainty was produced by the behaviour of analysed sources. We observed often a systematical trend in source location. Namely, the higher was the energy of the source the higher in the corona it was observed. Such systematical effect should be considered when estimating an actual source position. Thus, for a particular time interval we measured a spread of values obtained for different energies and the spread was assumed as a good approximation of uncertainty in source position. We usually obtained values about 5~arcsec. This systematical trend, higher the energy higher the altitude, is a common observational phenomenon \citep[e.g.][]{gallagher2002, sui2003, liu2008, sainthilaire2009}. Usually centroids of a source in energies below 10~keV are up to several arcsec below centroids in energies about 20 keV.

To estimate the altitude of an LTS we need to know its position in the {\it RHESSI} image and find the point on the solar photosphere above which the source is situated (reference point). The source position was defined, as mentioned, by the position of its centroid. The position of reference point was taken from: locations of foot point sources observed by {\it RHESSI} at impulsive phase (LDE1), locations of flare ribbons recorded by \textit{SoHO}/EIT (LDE2), method described by \citet{roy1975} (LDE3). The method allow to estimate heliographic coordinates for flares behind the solar limb. We plot a position (on the solar disk) of an active region in which an analysed behind-the-limb flare occurred as a function of time. The position is taken from positions of all on-disk flares which occurred in that active region. Then we extrapolate the position vs. time plot behind the limb to get a position of our behind-the-limb flare.

Since a moment of time for which position of the reference point was determined is different from times intervals analysed, all the positions were corrected to account for the solar rotation. Finally, the LTS altitude was calculated as the distance between its centroid and the reference point. The altitudes obtained were corrected for the projection effect. Errors in the location of reference point and the location of source centroid were included in the altitude errors. The relative error of altitude is about 10 percent.

The actual size of an LTS was determined using PIXON images reconstructed with the use of all detectors in which we observed the source at a given time. The size of the source  was estimated as the FWHM size. We did not observe important and systematical change of source size with energy. Thus, we decided to use PIXON images reconstructed in the energy interval $7-8$~keV for the estimation of a source size.

\subsection{Imaging spectroscopy of an LTS}

The main motivation for reconstructing images in narrow energy bands is the spectroscopy of the spatially resolved HXR sources. For this purpose we used OSPEX\footnote{http://hesperia.gsfc.nasa.gov/ssw/packages/spex/doc/ ospex\_explanation.htm} package. HXR continuum was fitted with thermal and, if it was observed, non-thermal components. We usually observed also line features at 6.7~keV and 8~keV \citep[]{phillips2006}. Physical parameters of observed sources were obtained from fits. Namely, for each time interval we obtained temperature and emission measure (thermal emission). If a non-thermal component was observed then we fitted it with double power-law function. We treated a break energy and power law index above this energy as free parameters. The power law index below the break energy was fixed at constant value equal to zero.

\subsection{Energy balance}

As mentioned in the Introduction, the presence of HXR emission from an LTS during the decay phase is evidence for energy release at that time. To calculate heating rate of an LTS we considered its energy balance during the decay phase. Three major cooling processes where included into this balance: expansion, radiation and conduction:

\begin{equation}
\label{e1}
\left (\frac{d{\mathcal E}}{dt}\right )_{obs} = \left( \frac{d{\mathcal E}}{dt} \right)_{ad} - E_C - E_R + E_H
\end{equation}

where: 

\begin{itemize}
	\item ${\mathcal E} = 3NkT$ is thermal energy density,
	\item $\left (\frac{d{\mathcal E}}{dt}\right )_{obs}$ is the decrease of ${\mathcal E}$ per second estimated from temperature ($T$) and density ($N$) values,
	\item $\left (\frac{d{\mathcal E}}{dt}\right )_{ad}$ is the decrease due to the adiabatic expansion of plasma in a source,
	\item $E_C$ is the energy loss due to thermal conduction,
	\item $E_R$ is the radiative loss, and
	\item $E_H$ is the heating rate or thermal energy release.
\end{itemize}

The values of $E_C$, $E_R$ and $E_H$ are in erg~cm$^{-3}$~s$^{-1}$. We calculated:

\begin{itemize}
	\item $\left (\frac{d{\mathcal E}}{dt}\right )_{ad} = 5kT\left (\frac{dN}{dt}\right)$,
	\item $E_C = 3.9\times10^{-7}T^{3.5}/(Lr)$ where $r$ is the LTS radius and $L$ is loop semi-length \citep[assuming that energy is conducted to the chromosphe,][]{jakimiec1997}, and
	\item $E_R = N^{2}\Phi(T)$ where $\Phi(T)$ is the radiative loss function taken from \citet[]{dere2009}.
\end{itemize}

We took the altitude of an LTS above the photosphere ($h$) as an approximation for $L$ in the expression for $E_C$. Of course, $h$ is smaller than $L$, but they do not differ too much. From Equation~(\ref{e1}) we calculated upper and lower limits for $E_H$. The upper limit, $(E_H)_{max}$, is calculated directly from Equation~(\ref{e1}) i.e. with $E_C$ described by Spitzer thermal conduction \citep[]{jakimiec1997}. However, it has been shown that the actual conductivity may be smaller than Spitzer conductivity \citep[]{luciani1983}. A lower (the lowest possible) limit, $(E_H)_{min}$, can therefore be obtained assuming $E_C = 0$, though this may not be physically realistic. The actual value of LTS heating rate must be contained between the upper and lower limit of $E_H$.

\subsection{Diagnostic diagram}

We used diagnostic (temperature-density, $T-N$) diagrams as an independent tool for analysing energy release. The method of investigating flare evolution by temperature-density diagrams was put forward by \citet[]{jakimiec1986, jakimiec1987, jakimiec1992} and \citet[]{sylwester1993}. It was shown that temporal evolution of flares does not depend much on their morphology. Thus, the diagrams allow us to study temporal changes of the energy release in flares qualitatively \citep[]{jakflud1992}. \citet[]{jakimiec1992} in their hydrodynamic modelling considered a single, symmetrical loop with constant cross-section and constant heating rate before a flare. The authors took the following variation of the heating rate with time $t$ in their calculations:

\begin{equation}
E_{H}(t) = \left\{ \begin{array}{ll}
A=const & \textrm{ for   } 0<t<t_{1}\\
Ae^{-\frac{t-t_{1}}{\tau}} & \textrm{ for   } t>t_{1}
\end{array} \right.
\label{energia}
\end{equation}
and studied two evolutionary tracks with $\tau=0$~s and $\tau=300$~s for a flaring loop of semi-length $L=2\times 10^{4}$ km. 

Non-flaring, steady loops having the same semi-length $L$ but different temperature $T$ and density $N$ lie along a straight line (the line of steady state loops, S--S line). Dependence between $T$ and $N$ for the S--S coronal loops is described by "scaling law" \citep[]{rosner1978}:

\begin{equation}
T=6.9 \times 10^{-4}(NL)^{1/2}
\end{equation}

\begin{figure}
\resizebox{\hsize}{!}{\includegraphics{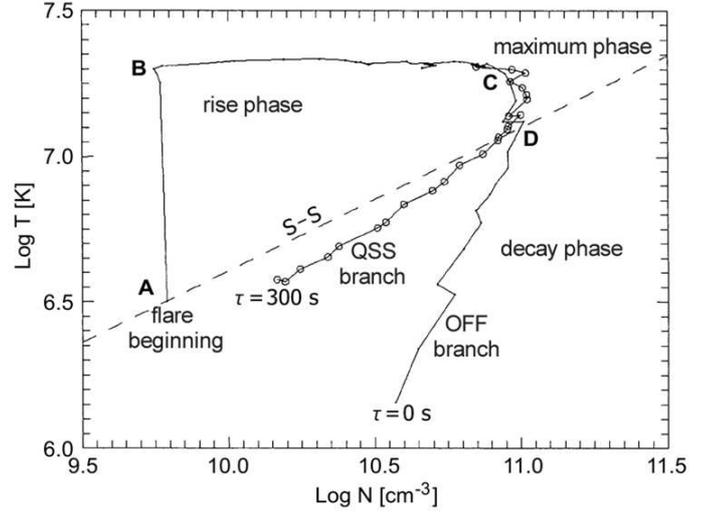}}
\caption{Density-temperature diagnostic diagram illustrating the dependence of the slope on the characteristic decay time 
of the heating rate ($\tau = 0$~s and $\tau = 300$~s cases are shown). The figure comes from the paper \citet []{jakimiec1992}.}
\label{diagnosticdiag}
\end{figure}

The inclination of the S--S line is $\xi=\frac{dlogT}{dlogN} \approx 0.5$. A flare evolution on the diagnostic diagram can be divided into few stages (see Fig.~\ref{diagnosticdiag}).

\begin{enumerate}
\item AB - Point A, lying on the S--S line, represents pre-flare conditions in a loop. Heating rate is constant, the loop is in steady-state. At the beginning of the flare energy is released abruptly causing increase of temperature (path AB).
\item BC - Temperature is constant, electron density increases gradually due to chromospheric evaporation. Rate of energy release starts to decreases at point C.
\item CD - Decreasing heating rate (point C) causes decrease of the temperature. However, temperature is still high enough to cause more chromospheric evaporation and increase of electron density (path CD). When released energy becomes insufficient the density in the loop begins to decrease (near the line S--S, point D).
\end {enumerate}

Evolution of a flare during the decay phase depends on how fast the heating rate $E_H$ decreases. In calculations two boundary scenarios observed during the decay phase may be present.

\begin{itemize}
\item If $E_H$ decreases very fast, with an e-folding decay time shorter than the thermodynamic decay time $\tau_{th} \propto L/\sqrt{T_0}$ \linebreak ($L$ - semilength of a loop, $T_0$ - initial temperature at the top of a flaring loop) then the flare evolves along the OFF branch ($E_H$ switched-off evolution). The evolution is characterized by a fast decrease of $T$ caused by the loop cooling due to thermal conduction losses and radiative losses not balanced by the heating rate, which is too low. This evolution is represented in Fig.~\ref{diagnosticdiag} by a branch for $E_H$ e-folding decay time $\tau = 0$~s.

\item If $E_H$ decreases slowly, with $\tau$ longer than $\tau_{th}$, the flaring loop will go through a sequence of steady-state configurations. In a $T-N$ diagram such a flare evolves along the QSS branch (quasi-steady-state evolution) which is parallel to the S-S line. During the QSS evolution $T$ drops slower than for the OFF evolution because the thermal conduction and radiative losses are balanced by the heating rate which is high enough. The QSS evolution is shown in Fig.~\ref{diagnosticdiag} by a branch for the $E_H$ e-folding decay time $\tau = 300$~s.
\end{itemize}
 
These results were obtained for a loop with a constant length. However, observations show that the height of flaring loops usually increases during a flare evolution. If  density also decreases (which is typical behaviour of solar flares) then QSS path will have inclination smaller then 0.5 \citep[see][]{kolomanski2007b}. In typical LDEs this altitude-density factor can lower the inclination of QSS path by about $0.1-0.2$.

The time behaviour of the energy release during the decay phase of the analysed flares was investigated using simplified diagnostic diagram (log T vs. log EM) obtained from the {\it GOES}/SEM data. Diagnostic diagrams obtained are shown in Figs.~\ref{fig20050730diag}, \ref{fig20050822diag} and \ref{fig20070125diag}.

\section{Results}

\subsection{The 30 July 2005 flare}

\begin{figure}
\resizebox{\hsize}{!}{\includegraphics{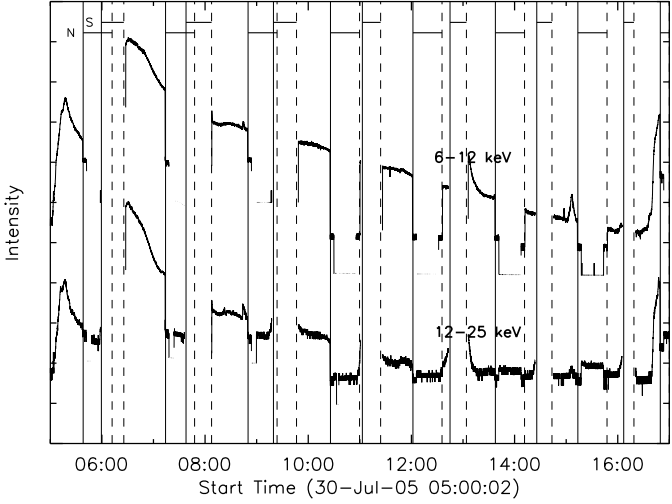}}
\caption{\textit{RHESSI} light curves for the 30 July 2005 flare. Vertical lines mark the boundaries of the satellite night (N) and the South Atlantic Anomaly (S) periods
(solid and dashed lines mark the beginning and the end of the periods, respectively). The light curves were shifted vertically for for clarity. The intensity is given in 
relative units.}
\label{fig20050730ligcur}
\end{figure}

{\it RHESSI} light curves of the LDE1 are shown in Fig.~\ref{fig20050730ligcur}. The flare emission could be analysed for eight intervals, given in Table~\ref{tab1} and \ref{tab2}. In the lower standard energy range $6-12$~keV, {\it RHESSI} recorded the emission for 10 hours during the decay phase until the beginning of the next flare. In the higher energy range $12-25$~keV the emission was detectable for 6 hours after flare maximum.

\begin{figure*}
\resizebox{\hsize}{!}{\includegraphics{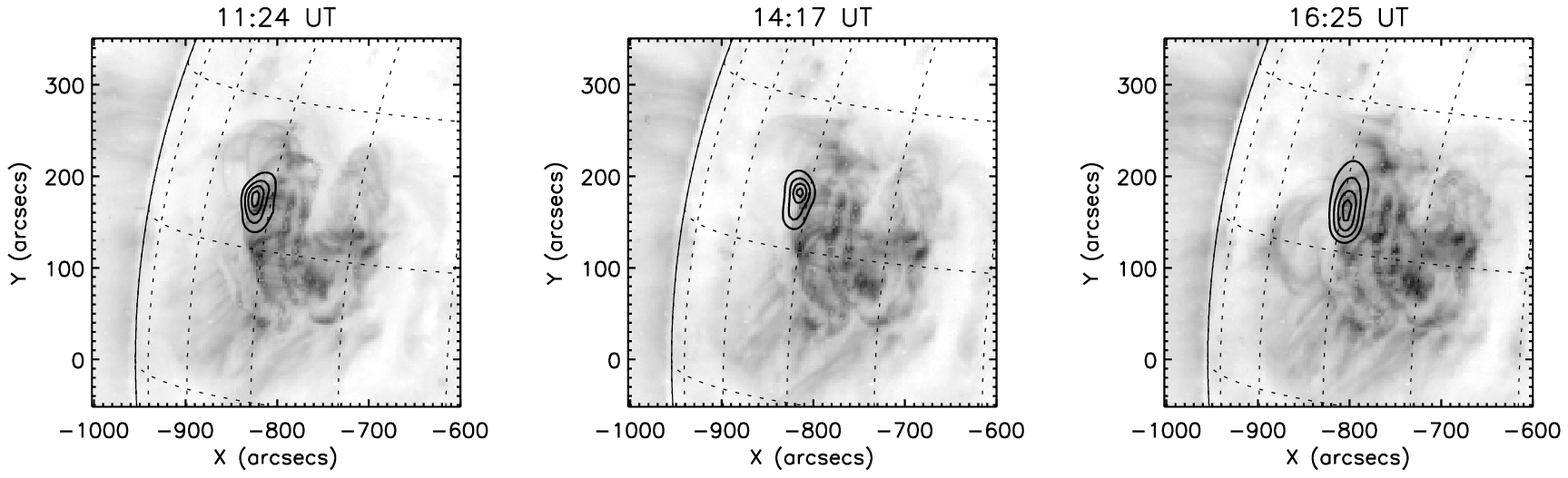}}
\caption{\textit{SoHO}/EIT 195~\AA \mbox{} images illustrating the decay phase of the 30 July 2005 flare. Contours show the emission in the $7-8$~keV range 
observed with \textit{RHESSI}.}
\label{fig20050730img}
\end{figure*}

Appearance of the flare and its evolution are shown in Fig.~\ref{fig20050730img}. The figure presents EIT~195~\AA \mbox{} images with a source observed by {\it RHESSI} overlaid on. As can be seen in EUV the LDE1 consisted of several loops forming an arcade. In X-rays there was one loop-top source located just above the arcade of loops  visible.

\begin{figure}
\resizebox{\hsize}{!}{\includegraphics{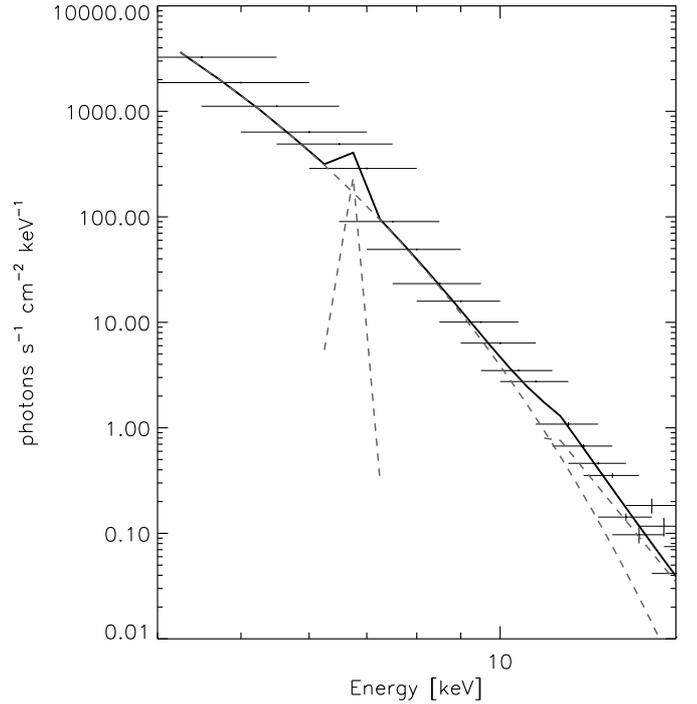}}
\caption{\textit{RHESSI} X-ray spectrum for the loop-top source of the 30 July 2005 flare recorded at 11:30~UT (horizontal bars, the bars widths corresponds to 
the energy bands widths). This spectrum was fitted using the thermal component, power-law non-thermal component and the spectral 
line complex at $6.7$~keV (gray dashed lines). The total fitted model is marked by black line.}
\label{fig20050730spect}
\end{figure}

Basic geometrical and physical properties of the loop-top source are given in Table~\ref{tab2}. The LTS size was about $32-35$~arcsec in diameter and it was not changing significantly with time. The LTS had an elliptical shape with longer axis running along the tops of the loops. The loops heights and LTS altitude slowly increased. The altitude grew from $47\times10^{3}$ to $60\times10^{3}\;{\rm km}$ with an average velocity of $1\;{\rm km}\:{\rm s}^{-1}$. The LTS spectrum could be fitted with one thermal component for all time intervals except for the three earliest ones for which we had to use a second component. The best fit was obtained for a thermal component with a power-law component (see Fig.~\ref{fig20050730spect}, see also Conclusions for details).

\begin{figure}
\resizebox{\hsize}{!}{\includegraphics{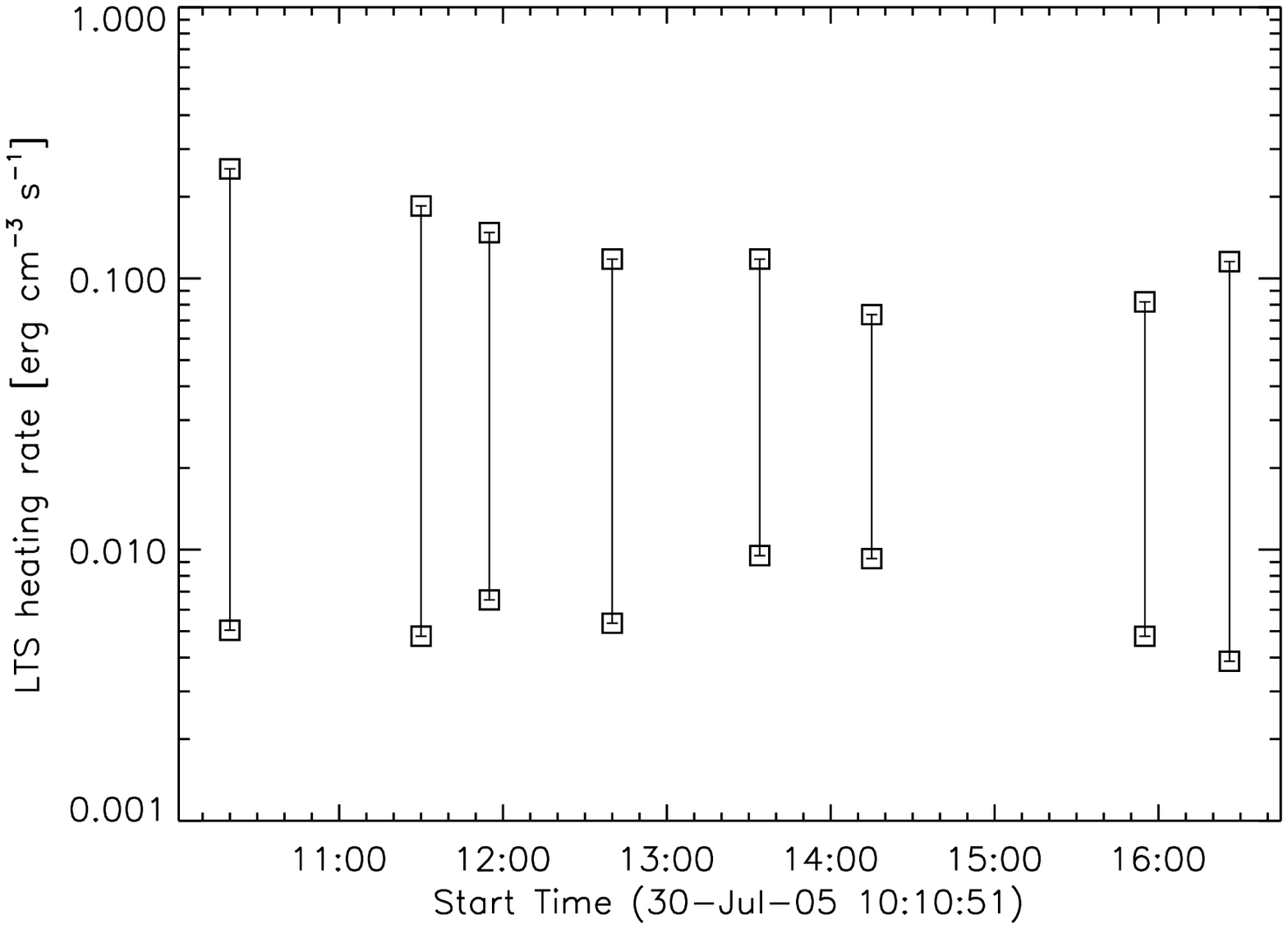}}
\caption{The 30 July 2005 flare -- estimated range of the heating rate of the loop-top source. The real value of the heating must be contained between upper (upper symbols) and lower 
(lower symbols) limit of heating rate.}
\label{fig20050730heat}
\end{figure}

\begin{figure}
\resizebox{\hsize}{!}{\includegraphics{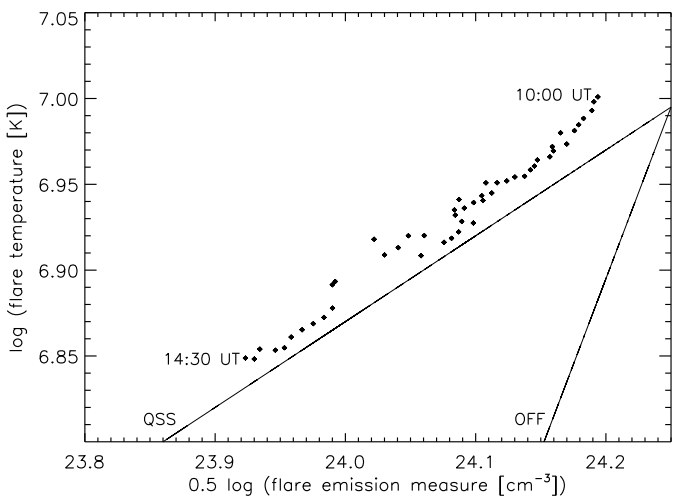}}
\caption{\textit{GOES} diagnostic diagram for the 30 July 2005 flare. The OFF and QSS paths of evolution are marked. The flare decay path runs along the QSS path. This confirms non-zero, slowly decreasing heating during decay phase of the flare.}
\label{fig20050730diag}
\end{figure}

Estimated values of LTS heating rate are shown in Fig.~\ref{fig20050730heat}. The actual value of the heating rate was about $0.1 - 0.01\;{\rm erg}\:{\rm cm}^{-3}\:{\rm s}^{-1}$. Non-zero value of $E_H$ is confirmed by diagnostic diagram (see Fig.~\ref{fig20050730diag}). The decay path of the flare runs along the QSS path i.e. there was continuous release of the energy and the rate of the release decreased very slowly which resulted in very slow cooling of LTS plasma. The characteristic time of temperature decrease was about 8 hours.

\begin{table}
\caption{Basic parameters of the loop-top source of the 30 July 2005 flare.} 
\label{tab2} 
\centering 
\begin{tabular}{c c c c c} 
\hline\hline 
\noalign{\medskip}
Time     & Size       & Altitude             & Temperature            & Emission                   \\ 
         & (Diameter) &                      &                        & Measure                     \\ 
$[$UT$]$ & [arcsec]   & [$10^{3}\;{\rm km}$] & [MK=$10^{6}\;{\rm K}$] & [$10^{47}\;{\rm cm}^{-3}$] \\ 
\noalign{\medskip}
\hline 
\noalign{\medskip}
10:20 & 42.2 & 47.5 & 11.2 & 9.90 \\ 
11:30 & 34.4 & 54.0 & 10.0 & 8.04 \\
11:55 & 34.8 & 52.9 & 9.3  & 9.01 \\
12:40 & 35.4 & 56.0 & 8.9  & 6.11 \\
13:34 & 24.4 & 58.0 & 8.0  & 6.85 \\
14:15 & 33.0 & 60.4 & 7.6  & 5.23 \\
15:55 & 32.0 & 59.5 & 7.9  & 1.31 \\
16:26 & 25.8 & 58.2 & 8.2  & 0.78 \\
\hline 
\end{tabular}
\end{table}

\subsection{The 22 August 2005 flare}

\begin{figure}
\resizebox{\hsize}{!}{\includegraphics{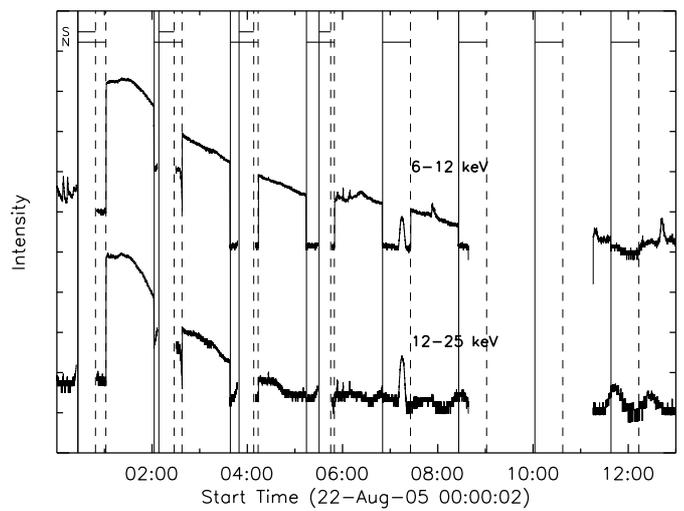}}
\caption{\textit{RHESSI} light curves for the 22 August 2005 flare. Vertical lines mark the boundaries of the satellite night (N) and the SAA (S) periods 
(solid and dashed lines mark the beginning and the end of the periods, respectively). The light curves were shifted vertically for for clarity. The intensity 
is given in relative units.}
\label{fig20050822ligcur}
\end{figure}

{\it RHESSI} light curves of the LDE2 are shown in Fig.~\ref{fig20050822ligcur}. The flare emission could be analysed for ten intervals (see Table~\ref{tab1} and ~\ref{tab3}). In the lower energy ranges $6-12$~keV, {\it RHESSI} recorded the emission for 7 hours during the decay phase until two-hour gap in the satellite data. In the higher range $12-25$~keV the emission was detectable for 3 hours after flare maximum.

\begin{figure*}
\resizebox{\hsize}{!}{\includegraphics{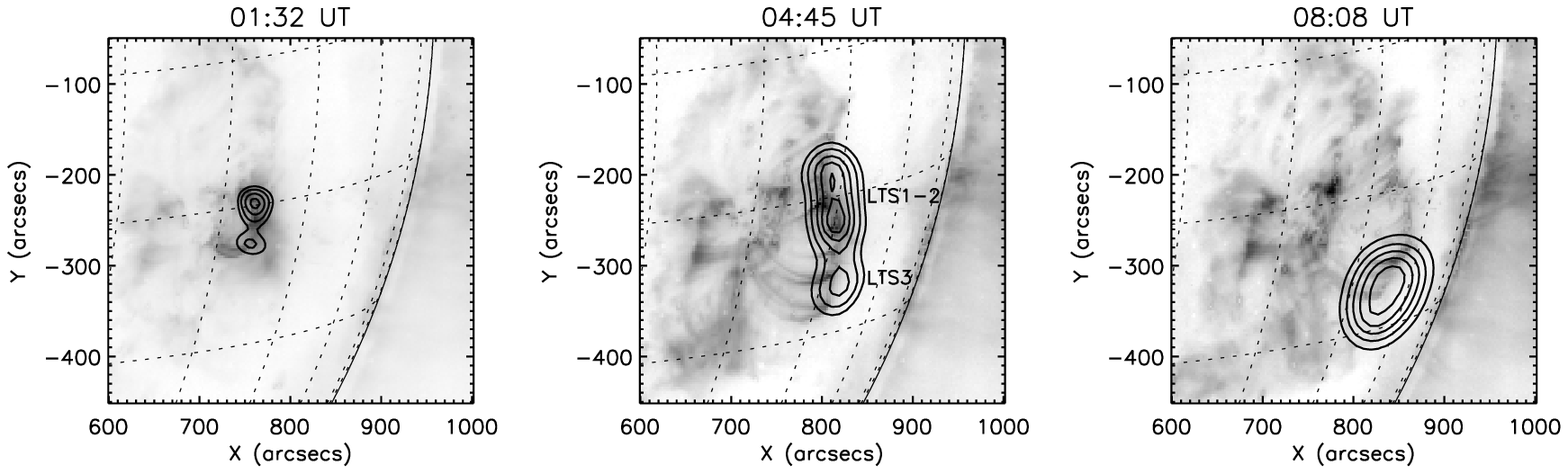}}
\caption{\textit{SoHO}/EIT 195~\AA \mbox{} images illustrating the decay phase of the 22 August 2005 flare. Contours show 
the emission in the $7-8$~keV range observed with \textit{RHESSI}.}
\label{fig20050822img}
\end{figure*}

Images of the flare and its evolution are shown in Fig.~\ref{fig20050822img} (EIT~195~\AA \mbox{} images with LTSs observed by {\it RHESSI} overlaid on). The LDE2 was visible as an arcade of loops in EUV while in X-rays there was three loop-top source located just above the arcade. The first two sources (LTS1, LTS2) were visible for about 4.5 hours until 6~UT. The third LTS appeared just before 6 UT and was present at least until the end of observations, i.e. for over two hours. The LTS1 and LTS2 were not separated well enough to be analysed individually, therefore we treated them as one source denoted as LTS1-2. 

Basic geometrical and physical properties of the loop-top sources are given in Table~\ref{tab3}. The LTS1-2 size grew from about 30~arcsec to 50~arcsec in diameter while LTS3 size remained almost unchanged (40~arcsec, excluding the last observation point). The three loop-top sources were located along the tops of the loops. The altitude of the LTS1-2 grew from $31\times10^{3}$ to $65\times10^{3}\;{\rm km}$ with an average velocity of $2\;{\rm km}\:{\rm s}^{-1}$. The LTS3 was located higher but is moved upwards with velocity two times lower. The emission of each LTS could be well described by a single thermal component for the whole decay phase.
  
\begin{figure}
\resizebox{\hsize}{!}{\includegraphics{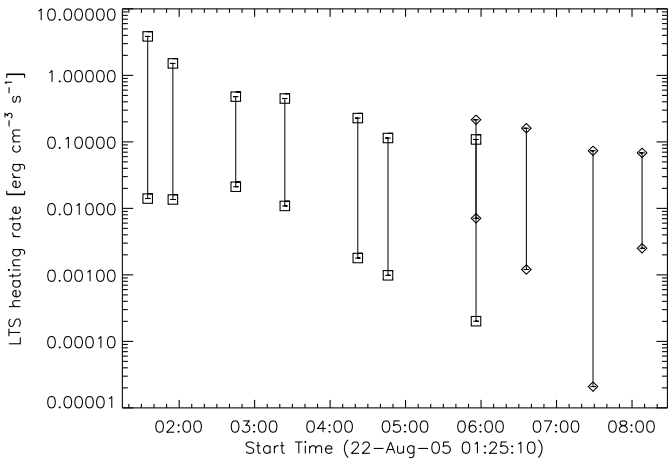}}
\caption{The 22 August 2005 flare  -- estimated range of the heating rate of the LTS1-2 source (squares) and LTS3 source (diamonds). The real value of the heating must be contained between upper 
(upper symbols) and lower (lower symbols) limit of heating rate.}
\label{fig20050822heat}
\end{figure}

\begin{figure}
\resizebox{\hsize}{!}{\includegraphics{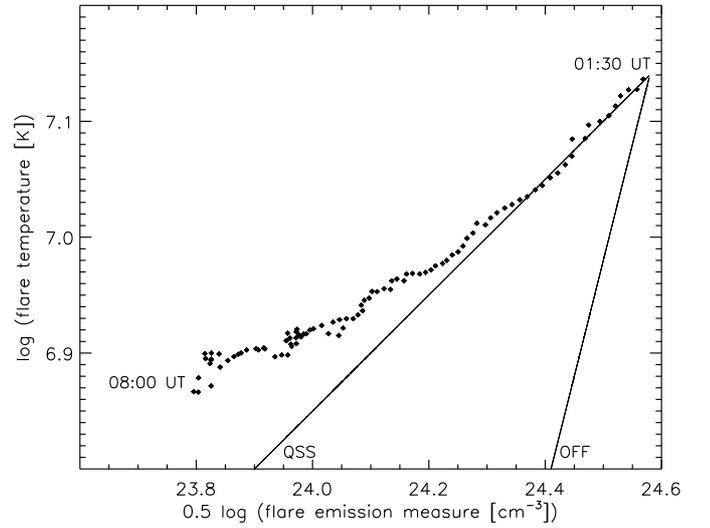}}
\caption{\textit{GOES} diagnostic diagram for the 22 August 2005 flare. The OFF and QSS paths of evolution are marked. The flare decay path 
runs along the QSS path (or even flatter). This confirms non-zero, slowly decreasing heating during decay phase of the flare.}
\label{fig20050822diag}
\end{figure}

The heating rate of LTS1-2 and LTS3 are shown in Fig.~\ref{fig20050822heat}. $E_H$ was non-zero and slowly decreased with time as can be seen. Such temporal behaviour of the heating rate is confirmed by diagnostic diagram (see Fig.~\ref{fig20050822diag}). The decay branch of the flare runs along the QSS path or even flatter (due to altitude-density factor). The characteristic time of temperature decrease was about 6 and 7 hours for LTS1-2 and LTS3, respectively.

\begin{table}
\caption{Basic parameters of the loop-top sources of the 22 August 2005 flare.} 
\label{tab3} 
\centering 
\begin{tabular}{c c c c c} 
\hline\hline 
\noalign{\medskip}
Time     & Size       & Altitude             & Temperature            & Emission                   \\ 
         & (Diameter) &                      &                        & Measure                     \\ 
$[$UT$]$ & [arcsec]   & [$10^{3}\;{\rm km}$] & [MK=$10^{6}\;{\rm K}$] & [$10^{47}\;{\rm cm}^{-3}$] \\ 
\noalign{\medskip}
\hline 
\noalign{\medskip}
LTS1-2&      &      &      &      \\
01:35 & 28.4 & 31.1 & 19.4 & 29.81 \\ 
01:55 & 31.8 & 33.4 & 15.6 & 35.12 \\
02:45 & 37.0 & 36.5 & 11.9 & 18.54 \\
03:24 & 37.2 & 42.6 & 12.3 & 7.26  \\
04:22 & 41.8 & 51.2 & 11.1 & 3.40  \\
04:46 & 53.6 & 57.4 & 10.1 & 3.20  \\
05:56 & 45.8 & 65.3 & 9.9  & 1.01  \\
\hline
\noalign{\medskip}
LTS3  &      &      &      &      \\
05:56 & 42.4 & 75.8 & 12.4 & 0.23 \\
06:36 & 39.0 & 78.0 & 11.1 & 0.82 \\
07:29 & 41.2 & 83.0 & 9.2  & 0.71 \\
08:08 & 62.8 & 78.6 & 9.9  & 0.85 \\
\hline 
\end{tabular}
\end{table}

\subsection{The 25 January 2007 flare}

\begin{figure}
\resizebox{\hsize}{!}{\includegraphics{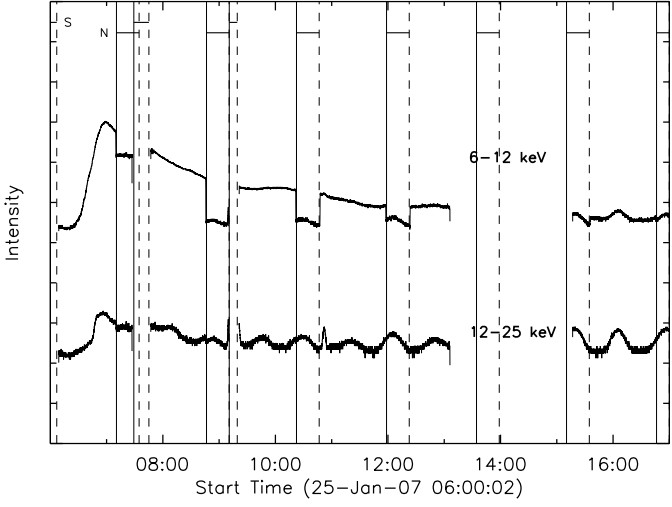}}
\caption{\textit{RHESSI} light curves for the 25 January 2007 flare. Vertical lines mark the boundaries of the satellite night (N) and the SAA (S) periods 
(solid and dashed lines mark the beginning and the end of the periods, respectively). The light curves were shifted vertically for for clarity. The intensity 
is given in relative units.}
\label{fig20070125ligcur}
\end{figure}

{\it RHESSI} light curves of the LDE3 are shown in Fig.~\ref{fig20070125ligcur}. The flare emission could be analysed for twelve intervals (see Table~\ref{tab1} and \ref{tab4}). In the energy range $6-12$~keV, {\it RHESSI} recorded the emission for 6 hours during the decay phase until two-hour gap in the satellite data. In the higher range $12-25$~keV the emission was detectable for 1 hour after the LDE3 maximum.

\begin{figure*}
\resizebox{\hsize}{!}{\includegraphics{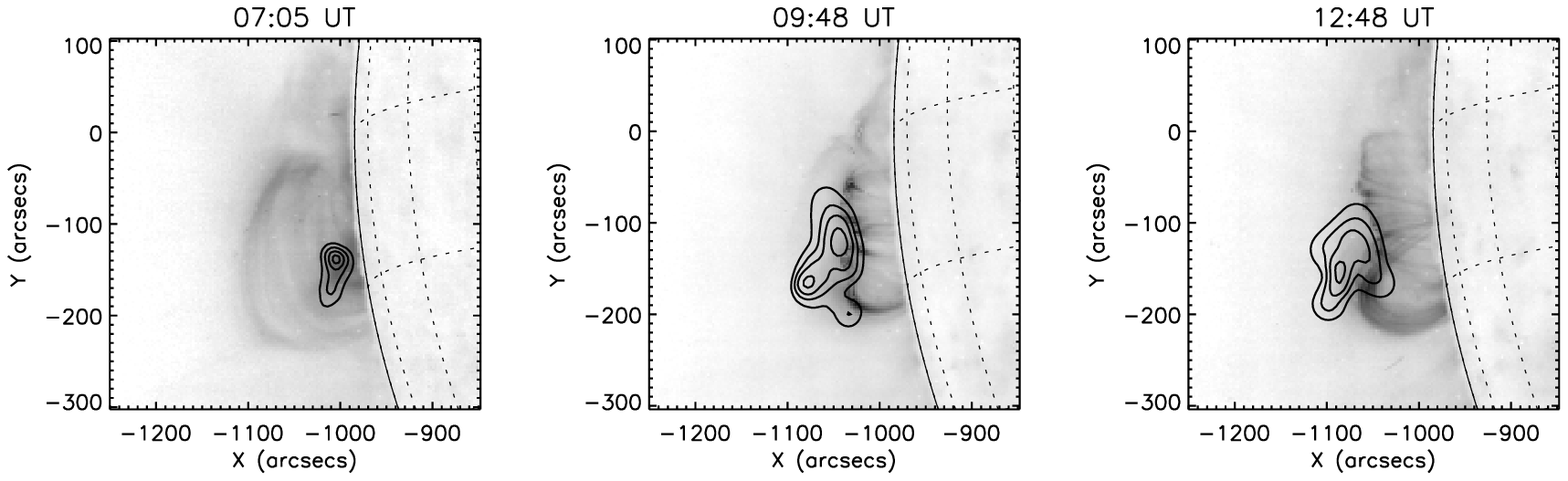}}
\caption{\textit{SoHO}/EIT 195~\AA \mbox{} images illustrating the decay phase of the 25 January 2007 flare. Contours show 
the emission in the $7-8$~keV range observed with \textit{RHESSI}.}
\label{fig20070125img}
\end{figure*}

Fig.~\ref{fig20070125img} shows images of the LDE3 during its decay phase (EIT~195~\AA \mbox{} images with intensity isolines from {\it RHESSI}). The LDE3 was a limb arcade flare with only footpoints obscured by the solar disk (see EIT images). X-rays emission came from a source located just above EUV loops. In some energies and time intervals there were two LTSs, but there was too few such observational points to analyse both LTSs separately and we decided to threat them as one source.

Basic geometrical and physical properties of the loop-top source are given in Table~\ref{tab4}. The LTS diameter changed from about 30~arcsec to 40~arcsec while its altitude grew from about $30\times10^{3}$ to $75\times10^{3}\;{\rm km}$ with an average velocity of $2\;{\rm km}\:{\rm s}^{-1}$. The emission of the LTS could be well described by a single thermal component for all time intervals.
  
\begin{figure}
\resizebox{\hsize}{!}{\includegraphics{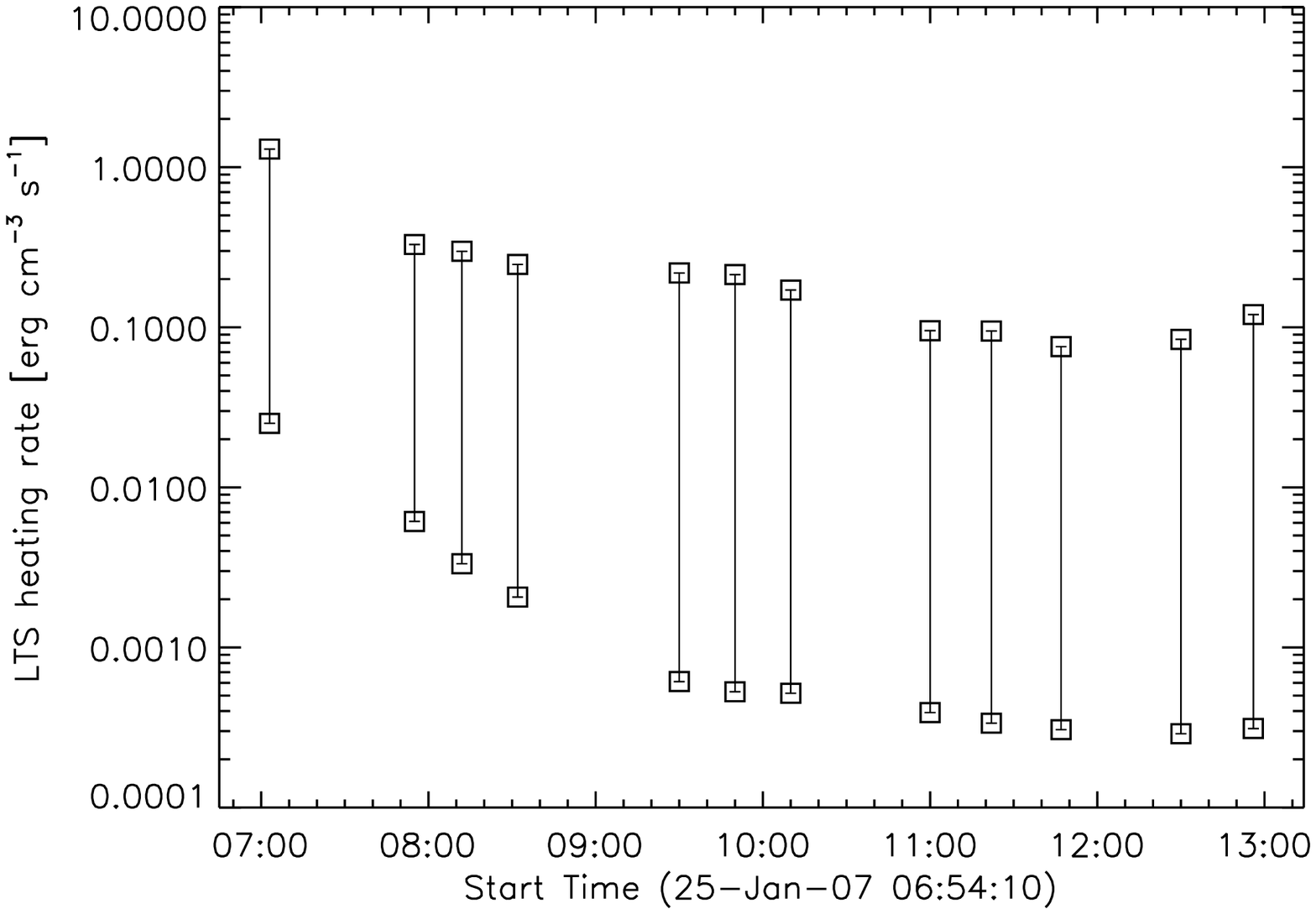}}
\caption{The 25 January 2007 flare -- estimated range of the heating rate of the loop-top source. The real value of the heating must be contained between upper 
(upper symbols) and lower (lower symbols) limit of heating rate.}
\label{fig20070125heat}
\end{figure}

\begin{figure}
\resizebox{\hsize}{!}{\includegraphics{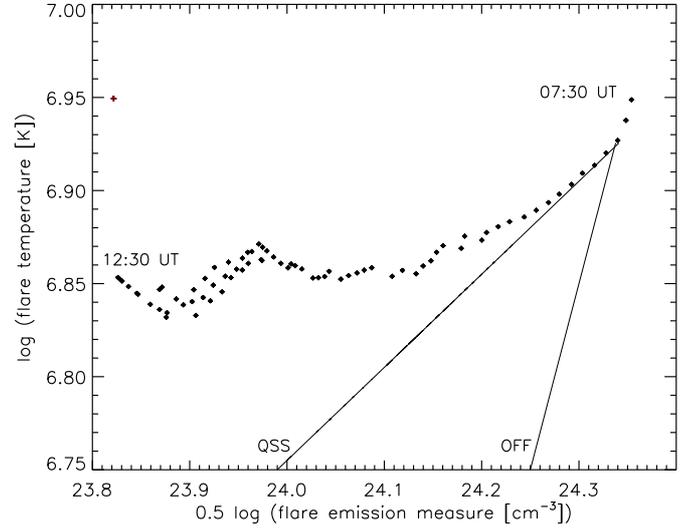}}
\caption{\textit{GOES} diagnostic diagram for the 25 January 2007 flare. The OFF and QSS paths of evolution are marked. The flare decay path 
runs along the QSS path (or even flatter). This confirms non-zero, slowly decreasing heating during decay phase of the flare.}
\label{fig20070125diag}
\end{figure}

The heating rate of the LTS of LDE3 are shown in Fig.~\ref{fig20070125heat}. $E_H$ was non-zero and slowly decreased with time. This temporal behaviour of the heating rate is confirmed by diagnostic diagram (see Fig.~\ref{fig20070125diag}). The decay branch of the flare runs along the QSS path or even flatter (due to altitude-density factor). The characteristic time of temperature decrease was very long i.e. 21 hours.

\begin{table}
\caption{Basic parameters of the loop-top source of the 25 January 2007 flare.} 
\label{tab4} 
\centering 
\begin{tabular}{c c c c c} 
\hline\hline 
\noalign{\medskip}
Time     & Size       & Altitude             & Temperature            & Emission                   \\ 
         & (Diameter) &                      &                        & Measure                     \\ 
$[$UT$]$ & [arcsec]   & [$10^{3}\;{\rm km}$] & [MK=$10^{6}\;{\rm K}$] & [$10^{47}\;{\rm cm}^{-3}$] \\ 
\noalign{\medskip}
\hline 
\noalign{\medskip}
07:03 & 22.4 & 30.5 & 13.3 & 13.49 \\ 
07:55 & 44.4 & 46.4 & 12.1 & 2.83  \\
08:12 & 42.4 & 44.8 & 11.5 & 1.52  \\
08:32 & 41.0 & 50.2 & 11.1 & 1.54  \\
09:30 & 40.6 & 51.0 & 10.8 & 0.75  \\
09:50 & 37.4 & 63.3 & 11.1 & 0.63  \\
10:10 & 47.6 & 55.6 & 10.8 & 0.74  \\
11:00 & 71.2 & 66.7 & 10.8 & 0.44  \\
11:22 & 55.0 & 68.6 & 10.1 & 0.42  \\
11:47 & 60.0 & 72.7 & 9.9  & 0.39  \\
12:30 & 53.0 & 74.1 & 9.9  & 0.36  \\
12:56 & 51.4 & 73.2 & 10.8 & 0.21  \\
\hline 
\end{tabular}
\end{table}

\section{Conclusions}

Previous results of the investigation of long-lasting HXR sources were based on the data giving rough information of physical parameters of a source. The {\it Yohkoh} HXT L channel gave an integrated flux in the wide energy band, $14-23$~keV. Using the {\it RHESSI} data we are now able to analyse the HXR flux and images with energy resolution as good as 1 keV. We analysed {\it RHESSI} images and spectral data for the three selected long-duration flares which occurred on 30 July 2005, 22 August 2005 and 25 January 2007.

The flux for energy above 12~keV was detected for 6, 3  and 1 hour during the decay phase of LDE1, LDE2 and LDE3, respectively. Below 12~keV emission was observed for at least 10, 7 and 6 hours after maximum of LDE1, LDE2 and LDE3, respectively. Such long-lasting high-energy emission is the first perceptible circumstantial evidence for the presence of heating processes during the decay phase.

For all analysed LDEs and all time intervals we observed large hard X-ray loop-top sources. Each LTS observed was smooth, without any internal structure despite high angular resolution of {\it RHESSI} (better than 10~arcsec). No LTS was visible in images reconstructed for a single narrower grid number $1-4$ (angular resolution $2.3-12$~arcsec), so a large LTS cannot be explained as a superposition of smaller unresolved sub-sources unless separation the sub-sources is smaller than angular resolution of the finest {\it RHESSI} grid ($2.3$~arcsec).

The emission of the LTSs can be described as a single thermal component for almost all time intervals. Temperature of the component was quite high (more than 10~MK in the early part of the decay phase) and decreased very slowly. A characteristic time of the decrease was at least 6 hours which is several times longer than a characteristic time of radiative cooling. A characteristic time of Spitzer conductive cooling is up to several minutes. This is indirect but strong evidence that efficient heating of LTSs is needed to enable their long-lasting existence.

This conclusion is confirmed by an energy balance calculation. The actual value of $E_H$ is larger than zero in all analysed time intervals, implying that the energy release occurred many hours after the flares maximum. Total amount of thermal energy released during the decay phase is big and is about $10^{31}\;{\rm erg}\:{\rm cm}^{-3}\:{\rm s}^{-1}$. This is not slight amount of energy and it can be comparable with or even greater than the amount of energy released during the impulsive phase of a flare \citep[see][]{jiang2006}. Non-zero, slowly decreasing heating during the decay phase was confirmed independently by the analysis of diagnostic diagrams. The decay evolution path runs along the QSS line or sometimes even flatter (due to altitude-density factor) for each of the three LDEs.

Similar results concerning the energy release during the decay phase of LDEs were obtained in analysis based on {\it Yohkoh}/SXT data \citep[e.g.][]{isobe2002, kolomanski2007b}. However, {\it RHESSI} data enable us to estimate much more reliable value of plasma temperature -- the SXT telescope had limited sensitivity to hot plasma. Hence values of temperature and therefore heating rate obtained from {\it RHESSI} data are higher than those obtained earlier from the SXT data. A proper solar flare model should be able to explain these more severe results.

In the case of LDE1 for the three earliest time intervals the emission of LTS cannot be described by a single spectral component regardless of the simple structure of the source. The analysis of spatially resolved spectra of the LDE1 showed that LTS emission at that time consisted of two qualitatively different spectral components. The best fit to the observed spectra was obtained for a thermal component plus a power-law component with index $\gamma = 9.7$ and break energy equal 12~keV. A slightly worse fit is obtained for two thermal components. This result puts some requirements on solar flare models regardless of the nature of the second component. If both components are thermal, some mechanism limiting fast mixing of hotter and cooler plasma in an LTS should exist. Moreover, the hotter component should be continuously heated by some mechanism. If one component is thermal and second is non-thermal, an acceleration process occurring within the hot (10 MK) and dense ($10^9-10^{10}\;{\rm cm}^{-3}$) region is needed. Both cases could be well explained by MHD turbulence in an LTS region. The turbulence produces large amount of small-scale reconnection regions where magnetic energy is continuously converted into thermal and non-thermal energy of plasma. Turbulence can also impede an escape of thermal energy generated by these small-scale reconnection regions outside since in coronal conditions turbulent conductivity may be several orders of magnitude lower than classical thermal conductivity parallel to magnetic field lines. The MHD turbulence inside LTSs were proposed by several authors \citep[e.g.][]{jakimiec1998, jiang2006, kontar2011}. A different explanation of two distinct thermal components observed within a flare coronal source was proposed by \citet{caspi2010}. Authors performed an analysis of short-duration flare and concluded that hotter ($>20$~MK) component came directly from coronal reconnection region while cooler component ($<20$~MK) originates mainly from chromospheric evaporation.

Here we reported three well observed events. In a future investigation we will analyse a larger number of flares. Such analysis will give us statistical information on efficiency of the energy release in LDEs.


\begin{acknowledgements}

The {\it RHESSI} satellite is NASA Small Explorer (SMEX) mission. We acknowledge many useful and inspiring discussions with Professor Micha\l \mbox{ }Tomczak. We also thank Barbara Cader-Sroka for editorial remarks unknown referee for useful comments and remarks. This investigation has been supported by a Polish Ministry of Science and High Education, grant No. N203 1937 33.

\end{acknowledgements}

\end{document}